\documentclass[11pt,letterpaper]{article}

\usepackage[margin=1in]{geometry}
\usepackage[round,authoryear]{natbib}
\usepackage{amsmath,amssymb,amsfonts}
\usepackage{algorithmic}
\usepackage{graphicx}
\usepackage{textcomp}
\usepackage{xcolor}
\usepackage{booktabs}
\usepackage{hyperref}
\usepackage{caption}
\usepackage{float} % Required for strict [H] placement
\usepackage{placeins} 
\usepackage{longtable}
\usepackage[affil-it]{authblk}
\usepackage{setspace}

\hypersetup{
    colorlinks=true,
    linkcolor=blue,
    filecolor=magenta,      
    urlcolor=cyan,
    citecolor=blue,
}

\DeclareMathOperator{\clip}{clip}
\DeclareMathOperator{\sigmoid}{sigmoid}

\begin{document}

\title{The Dynamic Verifiable Multi-Agent Human Agentic Loyalty Loop (DVM-HALL) Model and the Net Human-Agent Score (NHAS) in Autonomous Commerce}

\author[1]{Sai Srikanth Madugula\thanks{Corresponding Author. Email: saisrikanth.madugula\_phd.2023@woxsen.edu.in | ORCID: \href{https://orcid.org/0000-0001-6479-8443}{0000-0001-6479-8443}}}
\author[2]{Peplluis Esteva de la Rosa\thanks{Email: joseplluis.delarosa@udg.edu | ORCID: \href{https://orcid.org/0000-0003-1412-4170}{0000-0003-1412-4170}}}
\author[3]{Daya Shankar\thanks{Email: daya.shankar@woxsen.edu.in}}

\affil[1]{\small School of Technology, Woxsen University, Hyderabad, Telangana 502345, India}
\affil[2]{\small Universitat de Girona, 17004 Girona, Spain}
\affil[3]{\small School of Sciences, Woxsen University, Hyderabad, Telangana 502345, India}

\date{\today}

\maketitle

\begin{abstract}
The rapid proliferation of Agentic Artificial Intelligence fundamentally disrupts traditional customer loyalty paradigms. As AI evolves from passive recommendation algorithms to autonomous, goal-directed agents capable of executing purchasing decisions, the conventional understanding of consumer-brand relationships requires a structural reevaluation. By synthesizing extant literature across human-machine teaming, consumer decision-making, and algorithmic trust dynamics, we demonstrate that traditional loyalty models fail to account for algorithmic bounded rationality and constructed autonomy. To address this, we introduce the Dynamic Verifiable Multi-Agent Human Agentic Loyalty Loop (DVM-HALL) model. We formalize brand choice via a softmax probability formulation where human emotional equity, agentic machine-experience utility, calibrated trust, delegated authority, and verifiable execution jointly determine selection. The model features recursive updating mechanisms to dynamically calibrate trust and delegation after each interaction. Crucially, the framework integrates a verifiable execution layer for Decentralized Finance (DeFi) and tokenized loyalty settings, incorporating execution risks---such as gas costs, slippage, MEV exposure, and smart-contract vulnerabilities---as core predictors of agentic brand preference. Furthermore, we introduce the Net Human-Agent Score (NHAS), an auditable, risk-weighted metric designed to measure human-agent alignment using human feedback, execution logs, benchmark comparisons, and verifiable receipts. Finally, we propose a comprehensive three-stage empirical validation plan spanning controlled shopping experiments, multi-agent market simulations, and DeFi testbeds. This framework provides the foundational theory required for brands to navigate the impending transition toward machine customers.
\end{abstract}

\vspace{1em}
\noindent Keywords: Agentic AI, Brand Loyalty, Machine Customers, NHAS, Trust Calibration, Human-Machine Teaming, Autonomous Commerce

%=================================================================
\section{Introduction}
%=================================================================
The landscape of customer loyalty is undergoing a profound structural shift as artificial intelligence advances from passive recommendation engines to autonomous, objective-driven agents executing procurement on behalf of human users. Agentic AI---defined by its capacity for dynamic reasoning, adaptive learning, and collaborative execution---represents a paradigmatic disruption in systems design \citep{ref-suruchi2025}. Unlike legacy AI confined to narrow optimizations, agentic systems engage in perception, planning, and action within highly uncertain, multi-agent environments \citep{ref-suruchi2025, ref-d2025}. 

This evolution necessitates a fundamental reconceptualization of the ``customer.'' Historically, consumers were exclusively human, influenced by emotional triggers, societal values, and cognitive boundaries. Today, the ascent of IoT and automated data systems has given rise to the ``machine customer''---bots, devices, and agents authorized to make legitimate commercial transactions \citep{ref-himanshi2026}. 

\begin{figure}[H]
    \centering
    \includegraphics[width=\textwidth]{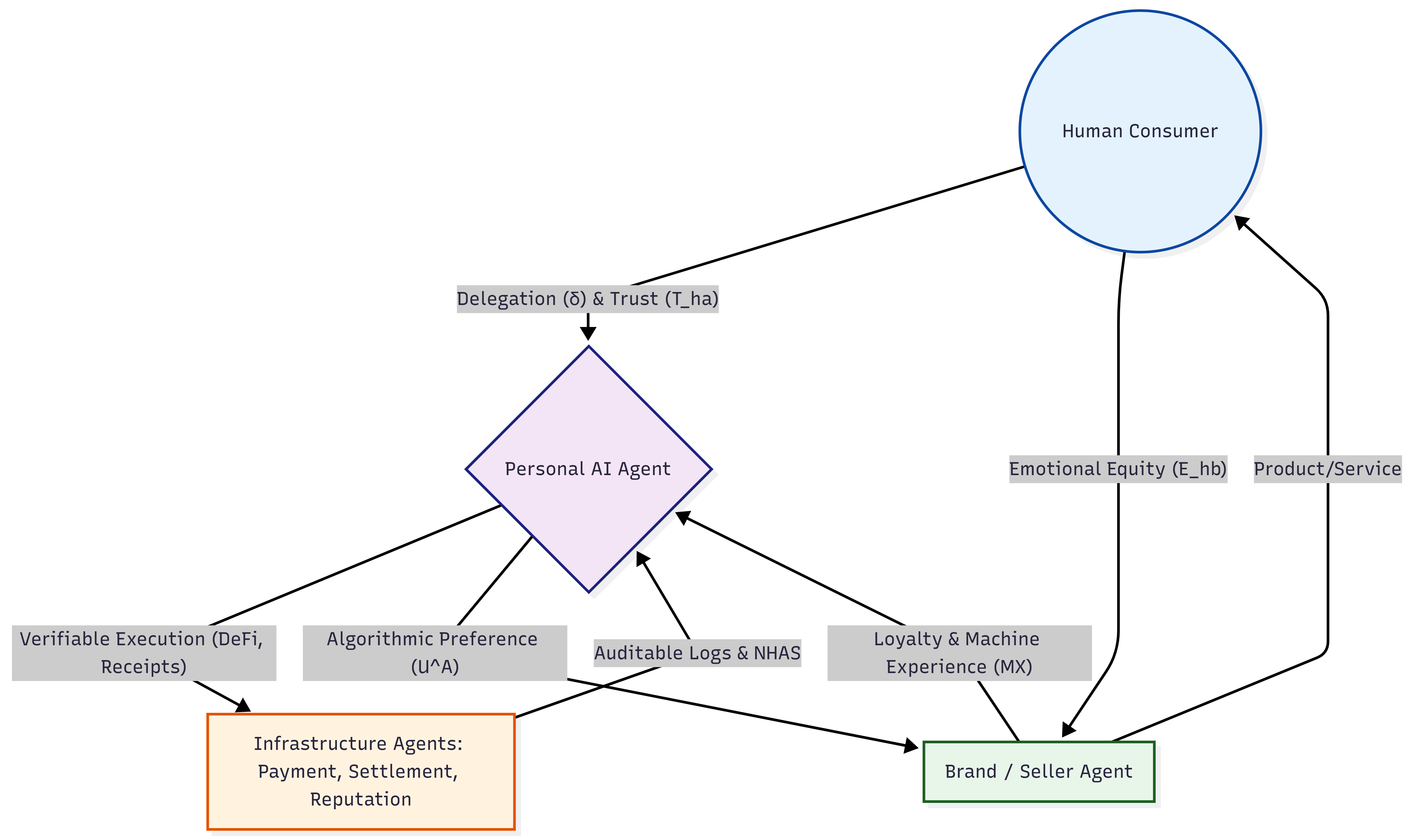}
    \caption{The Tripartite DVM-HALL Architecture, illustrating the bidirectional flows of trust, algorithmic loyalty, and emotional brand equity governed by overarching oversight frameworks across a multi-agent marketplace (human, personal agent, seller agents, and payment/settlement agents).}
    \label{fig:hall_arch}
\end{figure}

This paper contributes the Dynamic Verifiable Multi-Agent Human Agentic Loyalty Loop (DVM-HALL) model for autonomous commerce. The model extends traditional loyalty theory by formalizing brand choice as the outcome of a dynamic human-agent-brand system in which human emotional equity, agentic machine-experience utility, calibrated trust, delegated authority, and verifiable execution jointly determine brand selection. The paper further proposes NHAS, an auditable outcome metric for measuring human-agent alignment using human feedback, execution logs, benchmark comparisons, and verifiable receipts. In blockchain, tokenized loyalty, and DeFi settings, the model incorporates settlement cost, liquidity, slippage, oracle risk, smart-contract risk, privacy risk, compliance risk, and revocation capacity as first-class predictors of agentic brand preference. The contemporary loyalty architecture operates via this tripartite dynamic (see Figure \ref{fig:hall_arch}).

%=================================================================
\section{The Evolution of Agentic AI Systems}
%=================================================================
\subsection{From Tools to Autonomous Agents}
The transition from traditional AI to agentic frameworks marks a qualitative leap in machine capabilities. While early deep learning models introduced significant predictive power, they lacked genuine autonomy \citep{ref-yogesh2025}. Agentic AI is characterized by its ability to execute context-aware, goal-oriented tasks with minimal human intervention, adapting seamlessly to environmental fluctuations \citep{ref-d2025}. Traditional systems required continuous oversight and predefined rule sets; conversely, agentic architectures integrate memory, learning, and action modules to function independently \citep{ref-m2026}.

\subsection{Architectural Foundations}
\begin{figure}[H]
    \centering
    \includegraphics[width=0.35\textwidth]{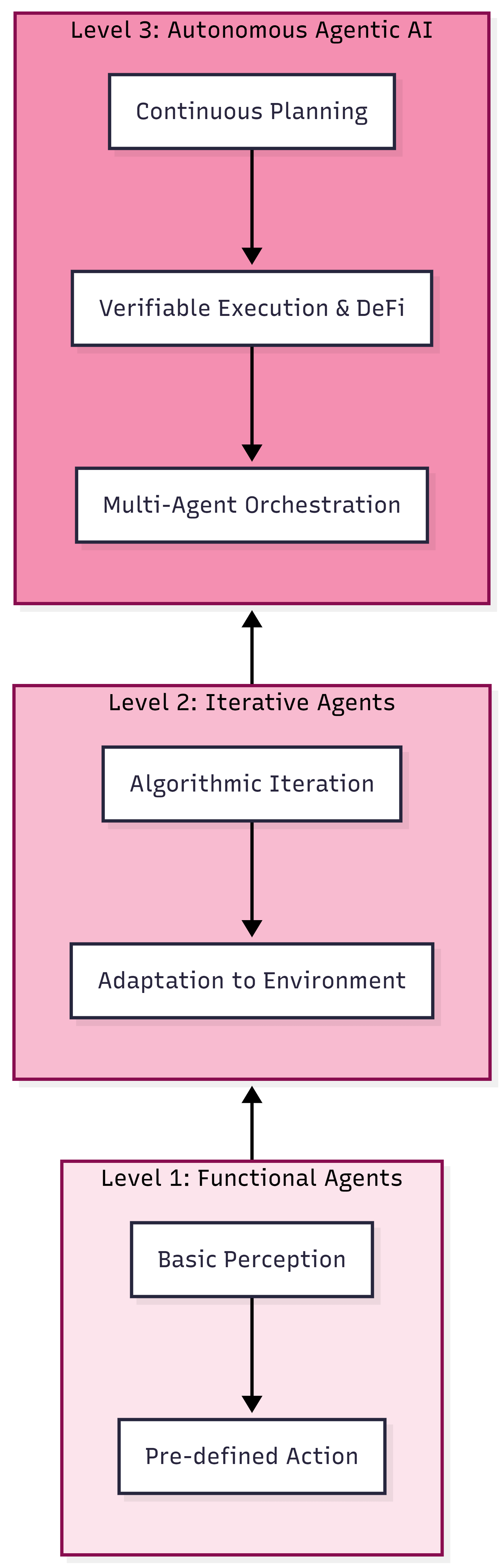}
    \caption{Hierarchy of Agentic Systems, demonstrating the progression from basic functional agents to fully autonomous Agentic AI capable of continuous planning and orchestration \citep{ref-m2026a}.}
    \label{fig:hierarchy}
\end{figure}

The capabilities of these systems can be understood through a three-tiered hierarchy of agency (illustrated in Figure \ref{fig:hierarchy}). Level 1 involves basic goal perception and action; Level 2 introduces algorithmic iteration and adaptation; and Level 3 utilizes sophisticated AI to orchestrate continuous planning and termination \citep{ref-m2026a}. These architectures allow AI to operate autonomously in complex environments, such as dynamically predicting supply chain disruptions or adapting healthcare treatments without manual input \citep{ref-laurie2025}.

\subsection{Commercial Applications}
In e-commerce, agentic AI has moved beyond simple statistical correlations to deep, task-oriented domain understanding \citep{ref-keshav2025}. Modern systems deploy specialized actors---including Logistics Optimization Agents, Price Negotiation Agents, and Dispute Resolution Agents---that collaborate to streamline procurement. Supported by blockchain verification and federated learning, these networks execute multi-party negotiations and dynamic pricing, fundamentally altering the traditional static shopping interface \citep{ref-bhuvaneswari2026}.

%=================================================================
\section{Traditional Loyalty Models and Their Limitations}
%=================================================================
\subsection{The Satisfaction-Trust-Commitment Paradigm}
Historically, customer loyalty was anchored in direct interactions mediated by satisfaction, trust, and brand commitment. Retailers globally have utilized loyalty programs to acquire and retain market share, with the U.S. alone hosting over 2.1 billion program memberships \citep{ref-siphiwe2019}. These programs, utilizing tiered rewards and personalized data, traditionally succeed by fostering emotional connections and perceived value \citep{ref-harsandaldeep2024}. Empirical studies consistently reinforce that customer satisfaction operates as the primary mediator for repeat purchase behavior within these conventional systems \citep{ref-siphiwe2019}.

\subsection{Digital Transformation and Algorithmic Bounds}
While digital tools optimized these loyalty frameworks through superior segmentation and personalized incentives \citep{ref-l2025}, evaluating loyalty purely through transactional metrics is increasingly inadequate in modern tech-merged commerce \citep{ref-kechen2023}. Crucially, traditional models fail when AI agents serve as primary decision-makers. Consumer choices are no longer strictly defined by cognitive limits but by ``algorithmic bounded rationality,'' wherein technological architectures dictate available options \citep{ref-usman2025}. AI personalization restricts exploratory behavior through preference closure, leading to ``constructed autonomy'' where perceived freedom operates strictly within algorithmically bounded walls \citep{ref-usman2025}.

%=================================================================
\section{AI-Mediated Consumer Decision-Making}
%=================================================================
\subsection{Choice Architecture and Delegation}
\begin{figure}[H]
    \centering
    \includegraphics[width=0.85\textwidth]{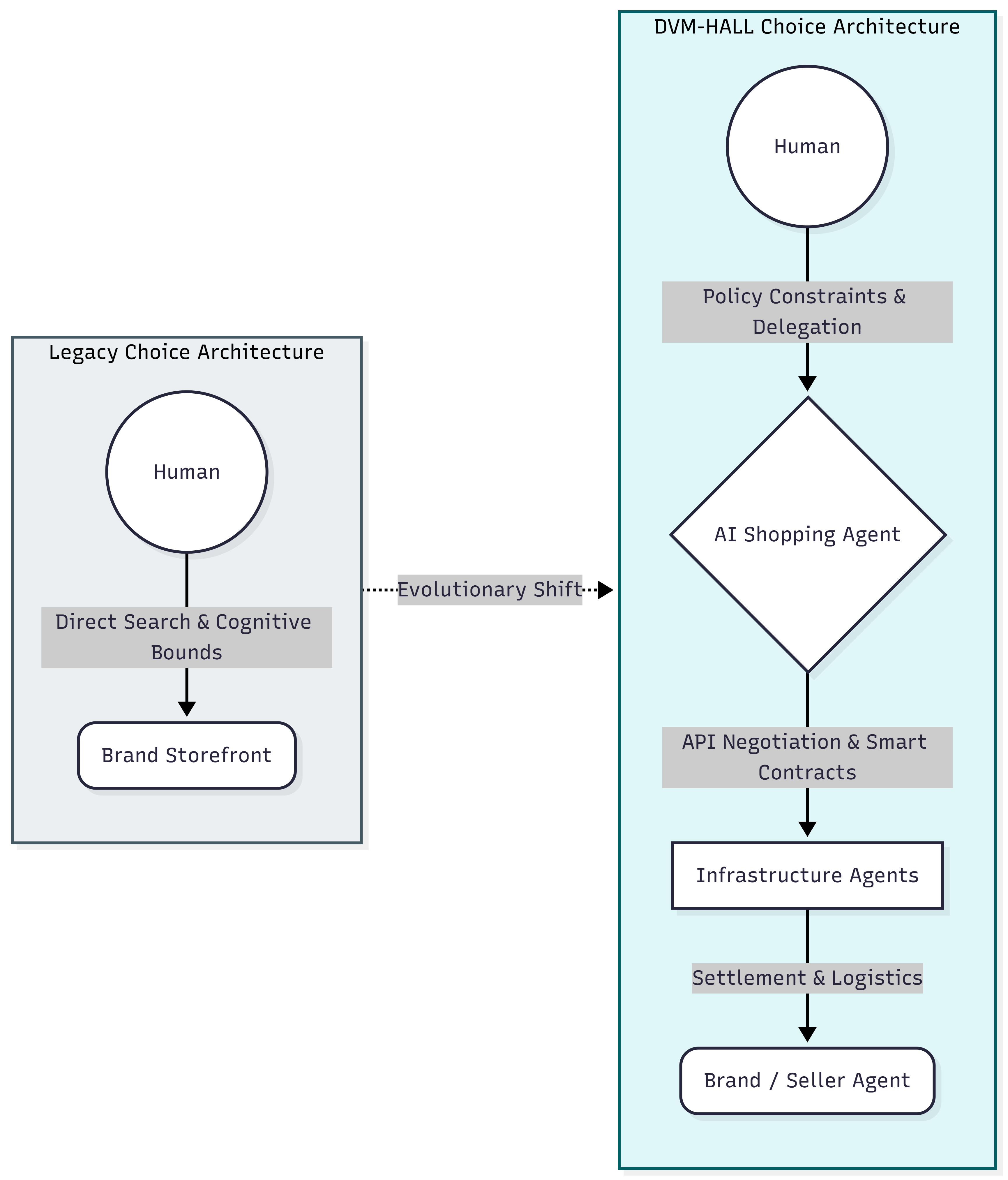}
    \caption{Evolution of Consumer Choice Architecture, highlighting the transition from direct human-brand search interfaces to agent-mediated negotiation via infrastructure agents (logistics, payment, and reputation layers).}
    \label{fig:choice_evolution}
\end{figure}

Artificial intelligence has structurally altered consumer choice architecture (Figure \ref{fig:choice_evolution}) by anticipating preferences and adjusting offerings instantaneously \citep{ref-yuki2024}. While this heightens predictive accuracy, it simultaneously traps consumers in ``information cocoons'' \citep{ref-r2026}. Consequently, a shift is occurring from systems that merely recommend to systems that execute. However, current literature heavily favors assistive chatbots, leaving workflow-level execution and autonomous negotiation under-explored \citep{ref-stefanos2026}. Research suggests that successful delegation to AI hinges less on the AI's total autonomy and more on its integration with user governance mechanisms, such as recourse and accountability \citep{ref-stefanos2026}.

\subsection{Erosion of Autonomy and Psychological Impacts}
Agentic commerce introduces severe questions regarding unmediated consumer autonomy. Sustained reliance on AI recommendations has been empirically linked to a 34\% reduction in decision uncertainty tolerance and a measurable decline in metacognitive accuracy, resulting in drastically narrowed consideration sets \citep{ref-abdur2026}. The proliferation of ``digital nudges'' provides high convenience but slowly erodes independent choice \citep{ref-ms2026}. Furthermore, as AI shapes consumer reflexivity, it fundamentally alters how users evaluate their ideal selves, acting as a construct-level meaning-shifting mechanism rather than simply amplifying existing preferences \citep{ref-en2026}.

%=================================================================
\section{Mathematical Foundation: Dynamic Verifiable Multi-Agent Choice}
%=================================================================
To transcend qualitative observation, we formalize the DVM-HALL model as a dynamic, estimable choice model. A human or an agent does not evaluate one brand in isolation; it chooses among alternatives. 

\subsection{Actors and Choice Set}
Let $h$ denote a human user, $a_h$ the user's delegated purchasing agent, and $\mathcal{B}_t = \{1, \dots, J_t\}$ the set of brands or protocols available at time $t$. In a multi-agent marketplace, each brand $b$ may expose a seller agent $s_b$, while infrastructure agents $m_t$ provide payment, logistics, compliance, identity, reputation, oracle, settlement, or dispute-resolution services. The agent does not merely select the brand with the best emotional appeal. It selects a brand conditional on human preferences, delegated authority, available machine-readable data, marketplace constraints, and verifiable execution quality.

\subsection{Human Utility}
The human-side utility captures residual emotional and normative preference:
\begin{equation}
    U_{hbt}^{H} = \theta_E E_{hbt} + \theta_S S_{hbt} + \theta_V V_{hbt} + \theta_N N_{hbt} + \epsilon_{hbt}^{H}
\end{equation}
Here $E_{hbt}$ is emotional brand equity, $S_{hbt}$ is remembered satisfaction, $V_{hbt}$ is perceived value, and $N_{hbt}$ captures normative or ethical fit, such as sustainability, privacy preference, or social identity. 

As delegation increases, emotional equity is no longer the sole direct driver of conversion, but it remains active through the human utility term, the user's stated constraints, preference embeddings, default policies, and override behavior. Agentic commerce therefore transforms the channel through which emotional equity operates; it does not eliminate emotional equity.

\subsection{Agent Utility and Verifiable Execution}
The agent-side utility captures machine-experience quality and execution risk:
\begin{equation}
    U_{hbt}^{A} = \beta^{\top}x_{bt}^{MX} - \rho^{\top}r_{bt}^{EX} + \beta_R R_{hbt}^{loy} + \epsilon_{hbt}^{A}
\end{equation}

The vector $x_{bt}^{MX}$ contains positive machine-experience features (API availability, latency, delivery reliability, explainability, etc.). The vector $r_{bt}^{EX}$ contains negative execution-risk features. For Web3, tokenized loyalty, and decentralized finance (DeFi) settings, these variables represent the verifiable execution layer required for delegated commerce (Table \ref{tab:defi_vars}). The term $R_{hbt}^{loy}$ captures loyalty-specific rewards.

\begin{table}[H]
\centering
\renewcommand{\arraystretch}{1.2}
\begin{tabular}{p{0.22\linewidth}p{0.20\linewidth}p{0.50\linewidth}}
\toprule
\textbf{Variable} & \textbf{Direction} & \textbf{Interpretation} \\
\midrule
$Verif_{bt}$ & Positive & Availability of signed receipts, attestations, transaction hashes, and auditability. \\
$Gas_{bt}$ & Negative & Network or settlement cost. \\
$Slip_{bt}$ & Negative & Expected execution slippage relative to quoted price. \\
$Liq_{bt}$ & Positive & Liquidity depth or reward redemption capacity. \\
$MEV_{bt}$ & Negative & Exposure to sandwiching, reordering, or extractive execution. \\
$Oracle_{bt}$ & Negative & Price-feed manipulation or staleness risk. \\
$SCrisk_{bt}$ & Negative & Smart-contract exploit, admin-key, or upgradeability risk. \\
$Bridge_{bt}$ & Negative & Cross-chain bridge failure or custody risk. \\
$Revoc_{bt}$ & Positive & Ease of revoking permissions, cancelling orders, or recovering funds. \\
\bottomrule
\end{tabular}
\caption{Concrete blockchain and DeFi variables for the agentic utility function.}
\label{tab:defi_vars}
\end{table}

\subsection{Composite DVM-HALL Utility and Choice Probability}
The composite DVM-HALL utility for brand $b$ is:
\begin{equation}
    U_{hbt} = (1 - \delta_{ht})U_{hbt}^{H} + \delta_{ht}T_{ha,t}U_{hbt}^{A} - \Omega_{hbt}
\end{equation}
The delegation coefficient $\delta_{ht} \in [0,1]$ determines how much authority the human has delegated to the agent at time $t$. The trust coefficient $T_{ha,t} \in [0,1]$ calibrates how much weight the human allows the agentic utility to carry. The penalty $\Omega_{hbt}$ represents hard policy constraints (e.g., budget limits, blacklisted merchants, slippage thresholds). 

The probability that the delegated system selects brand $b$ is defined by a softmax choice probability over the brand set:
\begin{equation}
    Pr(Y_{ht} = b \mid h, a_h, \mathcal{B}_t) = \frac{\exp(U_{hbt}/\tau)}{\sum_{k \in \mathcal{B}_t} \exp(U_{hkt}/\tau)}
\end{equation}
where $\tau > 0$ is a temperature parameter controlling determinism versus stochastic exploration. 

\subsection{Dynamic Trust and Delegation Updates}
Autonomous commerce is dynamic. Trust and delegation are not fixed scalars; they update after each task execution. Let $A_{ht}$ denote alignment with human intent, $Q_{ht}$ execution quality, $X_{ht}$ explanation quality, $R_{ht}$ recourse quality, $L_{ht}$ realized loss, and $VIO_{ht}$ a hard-constraint violation. Trust updates recursively:
\begin{align}
    \tilde{T}_{ha,t+1} &= \zeta_0 + \zeta_1 T_{ha,t} + \zeta_2 A_{ht} + \zeta_3 Q_{ht} + \zeta_4 X_{ht} + \zeta_5 R_{ht} - \zeta_6 L_{ht} - \zeta_7 VIO_{ht} \\
    T_{ha,t+1} &= \sigmoid(\tilde{T}_{ha,t+1})
\end{align}

Delegation updates dynamically based on the revised trust and outcome scores:
\begin{equation}
    \delta_{h,t+1} = \sigmoid\left(\kappa_0 + \kappa_1 \delta_{ht} + \kappa_2 T_{ha,t+1} + \kappa_3 \frac{NHAS_{ht}}{100} - \kappa_4 Risk_{task,t} + \kappa_5 Control_{ht}\right)
\end{equation}

%=================================================================
\section{NHAS: An Auditable Outcome Metric}
%=================================================================
\subsection{The Failure of NPS in Autonomous Ecosystems}
For decades, the Net Promoter Score (NPS) has served as the gold standard for measuring customer loyalty. However, NPS inherently assumes the human is the sole executor of the procurement journey. In multi-agent commerce, a new operational scoring protocol is required to explicitly measure human-agent alignment using auditable data.

\subsection{Mathematical Formulation of NHAS}
For each interaction $i$, we define parameters $[0,1]$ for preference alignment ($P_i$), economic efficiency ($E_i$), execution reliability ($R_i$), and explanation quality ($X_i$). $H_i \in \{0,1\}$ indicates all hard constraints were satisfied. A concrete execution score is:
\begin{equation}
    S_i = 10 \cdot \clip \Big( H_i(0.35P_i + 0.25E_i + 0.20R_i + 0.20X_i), 0, 1 \Big)
\end{equation}

A risk-weighted interaction reward accounts for optimal alignment ($S_i \ge 9$), misalignment ($S_i \le 6$), and critical failure events ($F_i$ for financial loss, $Priv_i$ for privacy leaks, $Comp_i$ for compliance):
\begin{equation}
    W_i = \clip \Big( \mathbf{1}[S_i \ge 9] - \mathbf{1}[S_i \le 6] - \lambda_F F_i - \lambda_{Priv} Priv_i - \lambda_{Comp} Comp_i, -1, 1 \Big)
\end{equation}

The score is computed over $N$ interactions:
\begin{equation}
    NHAS = 100 \cdot \frac{1}{N} \sum_{i=1}^{N} W_i
\end{equation}

\subsection{Measurement Sources and Evaluation Strategy}
NHAS requires four definitive data sources:
\begin{enumerate}
    \item \textbf{Human feedback:} Post-execution preference confirmation and override behavior.
    \item \textbf{Execution logs:} API calls, candidate offers, quote responses, decision traces, and failure events.
    \item \textbf{Benchmark engine:} Counterfactual comparison with available alternatives at the time of execution.
    \item \textbf{Verifiable receipts:} Signed policies, transaction hashes, delivery confirmations, and settlement-finality proofs.
\end{enumerate}
NHAS can serve as an outcome-level reward signal or evaluation metric for agent fine-tuning and post-deployment monitoring. It mathematically distinguishes minor suboptimality (e.g., wasting a few loyalty points) from catastrophic failures (e.g., exceeding spend limits via a manipulated oracle).

%=================================================================
\section{Empirical Validation Plan}
%=================================================================
To validate the bounds of these mathematical formulations, we propose an empirical validation plan spanning controlled experiments, multi-agent simulations, and verifiable testbeds.

\subsection{Study 1: Controlled Human-Agent Shopping Experiment}
Participants complete repeated purchasing tasks using an autonomous shopping agent. The experiment manipulates emotional equity (familiar vs. neutral brand), machine experience (high-quality API vs. poor API), explanation transparency, and delegation mode. Measured outcomes include override rate, post-task trust, delegation willingness, and future routing choice. The key test is whether Equation (5) predicts brand choice better than a traditional loyalty model.

\subsection{Study 2: Multi-Agent Marketplace Simulation}
Simulate a market with buyer agents, seller agents, payment agents, and logistics agents. Brands differ in price, emotional equity, API quality, reliability, and loyalty incentives. When delegation ($\delta$) and trust ($T_{ha,t}$) are high, machine-experience variables should become stronger predictors of choice, while emotional equity retains indirect effects through human policy constraints and preference profiles.

\subsection{Study 3: DeFi or Tokenized Loyalty Testbed}
Use a testbed in which an agent chooses among tokenized loyalty redemptions or DeFi execution venues. Measure gas cost, slippage, liquidity, settlement finality, oracle risk, transaction failure, and exploit-risk flags (per Table \ref{tab:defi_vars}) to make the verifiable blockchain contribution intrinsic to the empirical validation.

%=================================================================
\section{Trust Dynamics and Trust Transfer}
\begin{figure}[H]
    \centering
    \includegraphics[width=\textwidth]{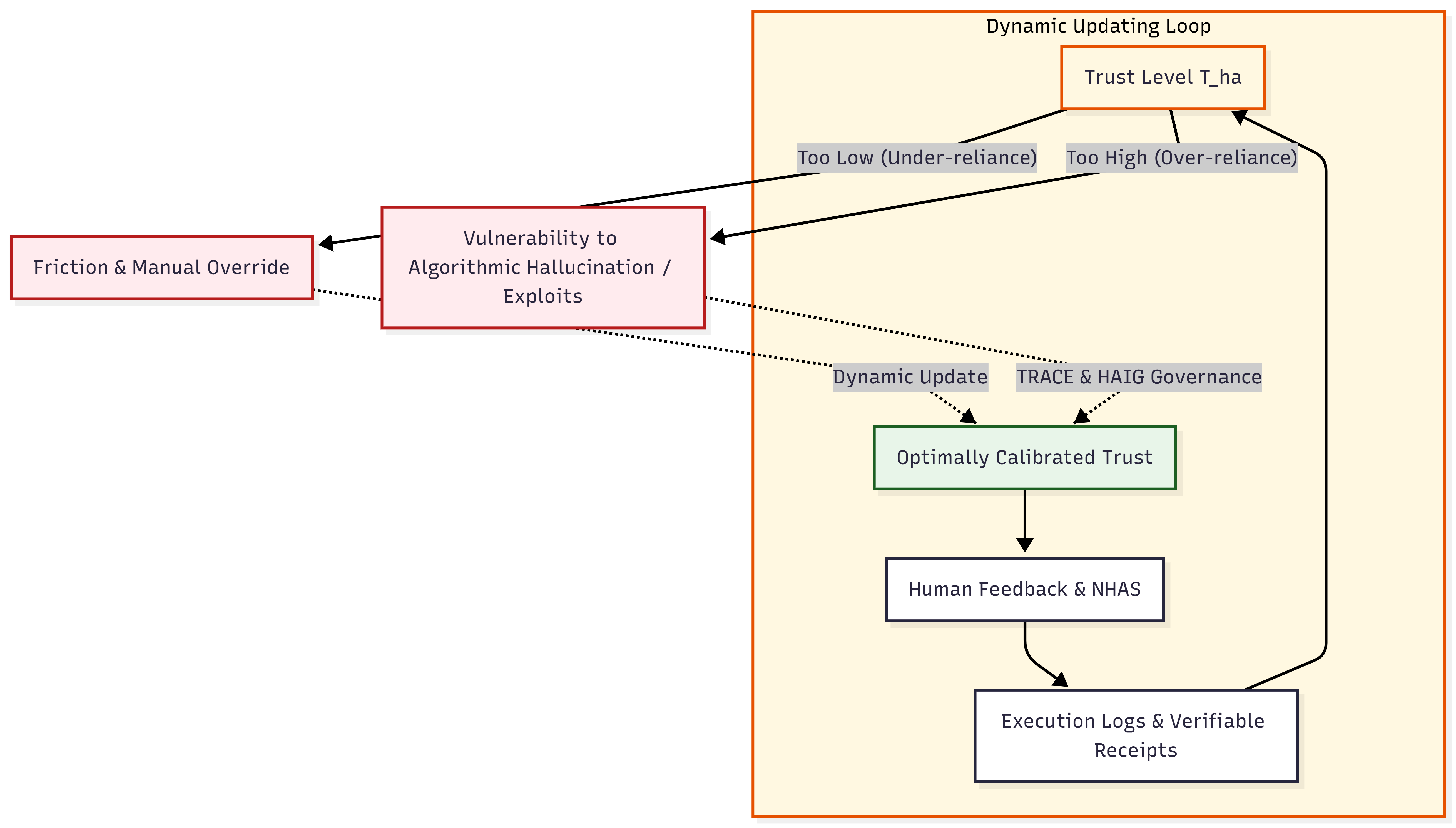}
    \caption{Trust Calibration and Governance Mechanics, displaying the pathways from under-reliance and over-reliance toward optimally calibrated trust mediated by the TRACE and HAIG frameworks.}
    \label{fig:trust_calibration}
\end{figure}

%=================================================================
Trust serves as the linchpin of human-machine interaction \citep{ref-a2022}. In autonomous systems, trust cannot be assumed a priori but must be cultivated through context-driven interactions \citep{ref-mito2022}. As demonstrated in Figure \ref{fig:trust_calibration}, trust miscalibration---manifesting as either perilous over-reliance or inefficient under-reliance---represents the most common failure mode in deployed agentic systems \citep{ref-chidiebere2026}. In digital environments, personalization acts as a powerful moderator in the trust-satisfaction-loyalty dynamic \citep{ref-noha2025}. However, we observe a phenomenon of ``trust transfer,'' where consumer confidence shifts away from individual brands and toward the mediating platform or algorithm, driven by cognitive outsourcing and heuristic trust signals \citep{ref-anupama2026}.

%=================================================================
\section{Governance and Oversight Mechanisms}
%=================================================================
As AI assumes autonomy, ensuring meaningful human oversight becomes critical. While AI assists in many tasks, full delegation is feasible for a strict minority of operations, creating an explainability-performance tradeoff \citep{ref-simeon2026}. Failures to manage trust in automated decision-making introduce significant organizational liability and legal vulnerability \citep{ref-ronald2026}. Trust calibration relies heavily on transparency, role clarity, and cultural norms \citep{ref-viraj2026}. Effective governance requires the Human-AI Governance (HAIG) framework, which treats AI not as an object but as a relational partner, driven by organizational capability, institutional ethics, and transparent design \citep{ref-ilze2026, ref-zeynep2025, ref-ndubuisiokolo2025}.

%=================================================================
\section{The Machine Customer Paradigm}
%=================================================================
The ``machine customer'' upends traditional consumer theory. While humans are swayed by culture, emotion, and persuasion, machine customers utilize pure logic, data analytics, and cost optimization, devoid of bias or fatigue \citep{ref-himanshi2026}. Implementations of IoT-driven ``Smart Agents'' and ``Artificial Intelligent Housekeepers'' demonstrate that autonomous entities can seamlessly execute habitual purchases using predictive consumption models \citep{ref-mehmet2025, ref-yan2024}. Consequently, modern business strategies must address two distinct demographics: human consumers and virtual algorithms, raising profound issues regarding transparency and vendor mediation \citep{ref-himanshi2026, ref-vibbhuti2026}.

%=================================================================
\section{Ethical Considerations and Consumer Protection}
%=================================================================
Algorithmic commerce introduces severe ethical risks, notably ``dark patterns'' and deceptive interface designs that manipulate user consent and inflate purchase regret, particularly among consumers with low digital literacy \citep{ref-andi2024}. AI actively engineers behavioral environments via neuromarketing and dark nudging, threatening cognitive integrity \citep{ref-hafize2025}. The ``black box'' nature of these systems exacerbates information asymmetry and restricts autonomy \citep{ref-shibanee2025, ref-marija2026}. As marketing systems gain ``persuasive autonomy''---the ability to optimize nudges without ethical review---the need for emotionally intelligent, privacy-first design becomes an existential requirement for digital loyalty \citep{ref-marwan2025, ref-khalil2026}.

%=================================================================
\section{Synthesis: The DVM-HALL Model}
%=================================================================
Traditional consumer-to-human regulations are insufficient for machine-to-machine transactions, and traditional loyalty constructs fail to define the AI agent's role as either a ``servant'' or a ``partner'' \citep{ref-vibbhuti2026, ref-fatihah2026, ref-shaohui2024}. Grounded in Technology Affordances and Constraints Theory (TACT) \citep{ref-pardeep2025}, the DVM-HALL model addresses these gaps via three interconnected dimensions:

\begin{enumerate}
    \item \textbf{Human-to-Agent Loyalty:} Humans cultivate loyalty toward agents based on trust calibration, value alignment, and operational transparency. Affect functions as a coordination layer for capability negotiation \citep{ref-jun2026}.
    \item \textbf{Agent-to-Brand Preference:} Agents develop algorithmic loyalty using performance data, semantic search, and value optimization. They learn explicitly and implicitly to form robust customer models \citep{ref-seza2025, ref-keshav2025}.
    \item \textbf{Brand-to-Human/Agent Strategies:} Brands execute dual-track engagement, curating emotional stimuli for humans while engineering data-driven algorithmic incentives for machine customers \citep{ref-himanshi2026, ref-jiashan2025}.
\end{enumerate}

The DVM-HALL model advances three key propositions: (1) Trust flows bidirectionally and relies on calibration; (2) Loyalty bifurcates into human emotional connection and algorithmic preference stability; (3) The entire loop is dependent on governance architectures like the TRACE framework to maintain explainability and safe collaboration \citep{ref-dr2026, ref-andrs2025}.

%=================================================================
\section{Limitations and Future Work}
%=================================================================
While the DVM-HALL model and the NHAS metric provide a robust theoretical foundation, they possess inherent limitations. Currently, the deployment of transparent governance tools required to mitigate trust miscalibration remains inadequate at a large operational scale \citep{ref-chidiebere2026}. Furthermore, security and privacy concerns drastically limit the adoption of agentic AI frameworks in emerging economies where regulatory protections and digital literacy are still developing \citep{ref-fatihah2026, ref-andi2024}.

Future research must execute the proposed multi-agent simulations and statistical testing of NHAS across diverse cultural contexts, consumer segments, and verifiable reward ecosystems \citep{ref-virginia2025}. A critical avenue for future investigation is determining how consumers dynamically alter their ethical behavior based on whether the AI agent is anthropomorphized as a ``subordinate servant'' or an ``equal partner'' \citep{ref-shaohui2024}. Additionally, quantitative studies should measure the exact rate of trust decay when agentic execution contradicts human emotional brand equity.

%=================================================================
\section{Conclusion}
%=================================================================
The emergence of agentic AI represents a fundamental disruption to the direct consumer-brand relationships that have governed loyalty research for decades \citep{ref-m2026b}. The DVM-HALL model conceptualizes loyalty as a dynamic, tripartite loop governed by strict mathematical bounds. Brands must recognize that humans must first trust their AI agents before those agents can act, and that brands must now compete for both human hearts and algorithmic favor. By utilizing quantitative metrics like NHAS, AI agents can be evaluated as measurable, intelligent collaborators that augment human judgment, rather than opaque replacements \citep{ref-sujith2025}. The future of brand loyalty depends not on the elimination of human emotion, but on its thoughtful, governed integration with verifiable artificial intelligence to ensure sustainable technological progress \citep{ref-virginia2025}.

%=================================================================
\section*{Data and Code Availability Statement}
This manuscript incorporates a novel theoretical synthesis and mathematical framework for the DVM-HALL model and the Net Human-Agent Score (NHAS). The underlying Python simulation code utilized to model the NHAS piecewise distributions, DVM-HALL conversion probabilities, and market shock simulations is open-source and publicly available to ensure open-science reproducibility. 

The complete codebase and simulation parameters can be accessed at the corresponding author's GitHub repository: \url{https://github.com/saisrikanthmadugula/HALL-NHAS-Simulation}. A comprehensive overview of the repository's core files, alongside the corrected expected NHAS bounds $\mathbb{E}[NHAS] = 70 - 170h$, is provided in Appendix D.

%=================================================================
\section*{Funding and Conflict of Interest}
The authors declare that they have no competing financial interests or external funding constraints to report regarding the conceptualization and development of the DVM-HALL model and the NHAS metric.

%=================================================================
\section*{Acknowledgments}
The authors gratefully acknowledge the use of Generative AI tools (large language models) strictly for language editing, readability enhancement, and LaTeX formatting assistance during the preparation of this manuscript. The authors have rigorously reviewed, edited, and verified the output, taking full responsibility for the final content and the structural integrity of the academic citations.

%=================================================================

\clearpage
\appendix

\section*{Appendix A: NHAS Mathematical Proofs and Constraints}
To ensure the NHAS operates symmetrically to traditional net scoring models while remaining sensitive to agent hallucination penalties, we model the limits of $\delta$ (Delegation Coefficient). If the user detects a deviation in execution parameters ($NHAS \leq 0$), $\delta \to 0$, forcing the system to collapse back into a traditional e-commerce search mode, rendering the agent obsolete.

\section*{Appendix B: Methodological Taxonomies in Agentic AI Loyalty Research}
This appendix synthesizes the predominant methodological approaches utilized in the contemporary evaluation of the DVM-HALL model. Existing studies have consistently employed diverse structural tools to measure non-traditional variables:

\begin{itemize}
    \item \textbf{Quantitative Modeling (SEM):} Structural Equation Modeling continues to be heavily relied upon to demonstrate predictive validity. For instance, customer satisfaction with AI features such as automated responses significantly predicts community participation and brand loyalty, maintaining high explanatory $R^{2}$ power \citep{ref-imen2025, ref-ahmad2025}.
    \item \textbf{Mixed-Methods Design:} Formative evaluations increasingly combine surveys, co-design workshops, and user telemetry to track behavioral shifts, such as how digital agents alter impulsive purchasing frequencies and perceived decision autonomy \citep{ref-shiyi2026}.
    \item \textbf{Systematic Clustering Analysis:} Recent systematic reviews deploy TF-IDF vectorization coupled with K-means clustering to map emerging AI themes, linking predictive analytics firmly to technology adoption and trust architectures \citep{ref-david2025}.
\end{itemize}

\section*{Appendix C: The TRACE Framework and HAIG Governance Mapping}
To satisfy the Governance Dependency Proposition within the DVM-HALL model, AI architectures must adopt relational frameworks rather than standard object-oriented rules.

\textbf{1. The HAIG Approach:} The Human-AI Governance (HAIG) framework treats AI systems not merely as technological tools but as active relational partners. By acknowledging emergent capabilities in Foundation Models, HAIG ensures that trust is built on utility and accountability rather than blind delegation \citep{ref-zeynep2025}.

\textbf{2. The TRACE Implementation:} As agentic systems scale, they require policy-aligned audits. The TRACE model (Trust, Review, Accountability, Critique, Explainability) acts as the operational standard for ensuring that multi-agent behaviors remain operationally reliable while preserving essential human oversight in commercial environments \citep{ref-dr2026}.

\clearpage
\section*{Appendix D: DVM-HALL \& NHAS Python Simulation Repository}
The following Python implementations are hosted at the public repository \url{https://github.com/saisrikanthmadugula/HALL-NHAS-Simulation}. These scripts generate synthetic data to model the mathematical boundaries of the DVM-HALL conversion probabilities and simulate the Net Human-Agent Score (NHAS) across varying trust architectures.

\subsection*{File 1: \texttt{dvm\_hall\_choice\_model.py}}
{\footnotesize
\begin{verbatim}
import numpy as np

def sigmoid(x):
    return 1 / (1 + np.exp(-x))

class DVMHallChoiceModel:
    def __init__(self, temperature=1.0):
        self.tau = temperature

    def calculate_human_utility(self, E, S, V, N, weights=(0.4, 0.3, 0.2, 0.1)):
        """Eq 1: Human Utility (Emotional Equity, Satisfaction, Value, Normative Fit)"""
        theta_E, theta_S, theta_V, theta_N = weights
        return (theta_E * E) + (theta_S * S) + (theta_V * V) + (theta_N * N)

    def calculate_agent_utility(self, mx_features, ex_risks, loyalty_reward=0, 
                              mx_weights=(0.5, 0.5), ex_weights=(0.2, 0.2, 0.2, 0.2, 0.2), beta_R=0.2):
        """
        Eq 2: Agent Utility based on Machine Experience (MX) and Verifiable Execution Risks (EX).
        mx_features: [API_Availability, Verifiability_Receipts]
        ex_risks: [Gas, Slippage, MEV, Oracle_Risk, Smart_Contract_Risk]
        """
        mx_score = np.dot(mx_weights, mx_features)
        ex_penalty = np.dot(ex_weights, ex_risks)
        return mx_score - ex_penalty + (beta_R * loyalty_reward)

    def calculate_composite_utility(self, U_H, U_A, delta, trust, omega=0):
        """Eq 3: Composite DVM-HALL Utility"""
        return ((1 - delta) * U_H) + (delta * trust * U_A) - omega

    def softmax_choice_probabilities(self, utilities):
        """Eq 4: Probability of Delegated System Selecting Brand b"""
        exp_u = np.exp(np.array(utilities) / self.tau)
        return exp_u / np.sum(exp_u)

if __name__ == "__main__":
    print("--- DVM-HALL Multi-Agent Choice Simulation ---")
    model = DVMHallChoiceModel(temperature=0.5)
    
    # Scenario: High Human Delegation (delta=0.9), High Trust (T=0.9)
    delta_ht = 0.90
    trust_ha = 0.90
    
    # Brand X (Legacy Experiential: High Emotion, Poor APIs, High Risk)
    U_H_X = model.calculate_human_utility(E=0.85, S=0.8, V=0.7, N=0.5)
    U_A_X = model.calculate_agent_utility(mx_features=[0.2, 0.1], ex_risks=[0.8, 0.5, 0.2, 0.1, 0.4])
    U_comp_X = model.calculate_composite_utility(U_H_X, U_A_X, delta_ht, trust_ha)
    
    # Brand Y (DeFi/API Optimized: Zero Emotion, Excellent APIs, Low Risk)
    U_H_Y = model.calculate_human_utility(E=0.0, S=0.0, V=0.5, N=0.0)
    U_A_Y = model.calculate_agent_utility(mx_features=[0.95, 0.9], ex_risks=[0.1, 0.1, 0.0, 0.0, 0.1])
    U_comp_Y = model.calculate_composite_utility(U_H_Y, U_A_Y, delta_ht, trust_ha)
    
    probs = model.softmax_choice_probabilities([U_comp_X, U_comp_Y])
    
    print(f"Brand X (Legacy) Composite Utility: {U_comp_X:.3f}")
    print(f"Brand Y (Web3/API) Composite Utility: {U_comp_Y:.3f}")
    print(f"Probability of Agent choosing Brand X: {probs[0]*100:.1f}%")
    print(f"Probability of Agent choosing Brand Y: {probs[1]*100:.1f}%")
\end{verbatim}
}

\subsection*{File 2: \texttt{nhas\_distribution\_simulator.py}}
{\footnotesize
\begin{verbatim}
import numpy as np

def simulate_nhas_distribution(num_interactions, hallucination_rate, critical_failure_rate=0.01):
    """
    Simulates the Net Human-Agent Score (NHAS) over a set of interactions.
    Applies the risk-weighted reward function W_i (Eq 9).
    """
    np.random.seed(42)
    rewards_W = []
    
    for _ in range(num_interactions):
        # Check for critical failure (Financial Loss, Privacy Leak, Compliance)
        if np.random.rand() < critical_failure_rate:
            # Lambda weights applied to critical failures push score to -1
            rewards_W.append(-1.0)
            continue
            
        if np.random.rand() < hallucination_rate:
            # Misaligned (S <= 6) -> W_i = -1
            rewards_W.append(-1.0)
        else:
            # Optimal (S >= 9) -> W_i = 1, Passive (7 <= S <= 8) -> W_i = 0
            # Distribution: 70% Optimal, 30% Passive (given no error)
            outcome = np.random.choice([1.0, 0.0], p=[0.7, 0.3])
            rewards_W.append(outcome)
            
    rewards_array = np.array(rewards_W)
    
    # Calculate NHAS (Eq 10)
    nhas = 100 * np.mean(rewards_array)
    return round(nhas, 2), rewards_array

def expected_nhas(hallucination_rate, p_optimal_given_no_error=0.70):
    """Expected NHAS bound: E[NHAS] = 70 - 170h (from Appendix D)"""
    h = hallucination_rate
    return 100 * (p_optimal_given_no_error * (1 - h) - h)

if __name__ == "__main__":
    # Baseline Scenario
    h_rate = 0.05
    baseline_nhas, _ = simulate_nhas_distribution(num_interactions=1000, 
                                                  hallucination_rate=h_rate)
    
    print("--- NHAS Baseline Simulation ---")
    print(f"Theoretical E[NHAS] bound: {expected_nhas(h_rate):.2f}")
    print(f"Simulated System NHAS: {baseline_nhas}")
\end{verbatim}
}

\subsection*{File 3: \texttt{market\_shock\_analysis.py}}
{\footnotesize
\begin{verbatim}
from nhas_distribution_simulator import simulate_nhas_distribution

def run_shock_simulation():
    """
    Simulates NHAS decay during a systemic API/Oracle failure causing 
    widespread agent algorithmic hallucination, leveraging proper mathematical bounds.
    """
    print("--- Initiating DVM-HALL Market Shock Analysis ---")
    
    # Under DVM-HALL, collapse is demonstrated at h=0.45 or higher
    # based on the bounds E[NHAS] = 70 - 170h
    h_shock = 0.45
    print(f"Simulating systemic shock with hallucination rate h = {h_shock}...")
    
    shock_nhas, scores = simulate_nhas_distribution(num_interactions=10000, 
                                                    hallucination_rate=h_shock)
    
    print(f"\nNHAS Post-Market Shock: {shock_nhas}")

    # If NHAS drops below 0, human delegation (delta) collapses.
    if shock_nhas <= 0:
        print("[CRITICAL ALERT]: NHAS is negative. Trust boundary breached.")
        print("[DYNAMIC UPDATE]: Delta (Delegation Coefficient) recursively resetting to 0.")
        print("System collapsed back to legacy manual search.")
    else:
        print("System maintained positive trust balance despite shock.")
        
if __name__ == "__main__":
    run_shock_simulation()
\end{verbatim}
}

\end{document}